# The role of surface water in the geometry of Mars' valley networks and its climatic implications

H.J. Seybold[1]*, E. Kite[2], J.W. Kirchner[1,3]

[1]ETH Zurich, 8092 Zurich, Switzerland
[2]University of Chicago, Chicago, IL 60637, USA
[3]Swiss Federal Research Institute WSL, 8903 Birmensdorf, Switzerland

*Correspondence to: hseybold@ethz.ch

**Abstract**:

Mars' surface bears the imprint of valley networks formed billions of years ago and their relicts can still be observed today. However, whether these networks were formed by groundwater sapping, ice melt, or fluvial runoff has been continuously debated. These different scenarios have profoundly different implications for Mars' climatic history, and thus for its habitability in the distant past. Recent studies on Earth revealed that channel networks in arid landscapes with more surface runoff branch at narrower angles, while in humid environments with more groundwater flow, branching angles are much wider. We find that valley networks on Mars generally tend to branch at narrow angles similar to those found in arid landscapes on Earth. This result supports the inference that Mars once had an active hydrologic cycle and that Mars' valley networks were formed primarily by overland flow erosion with groundwater seepage playing only a minor role.

**One Sentence Summary:** Valley network branching angles suggest that Mars' ancient hydrology was not dominated by groundwater flow.

Decades of satellite missions to Mars have shaped an evolving narrative of the history of water on the red planet. Early missions returned images showing ancient channel networks but no concrete evidence for flowing water (*1-4*). Mars' cold present-day climate, combined with Earth-analogue fieldwork, led to the hypothesis that Mars' channel networks could have been carved by streams sourced from groundwater springs (*5-8*). However, it remains unclear how these processes could lead to branching networks of kilometer-wide valleys incised into bedrock (*9, 10*). Uncertainty concerning the valley incision process is closely coupled to the question of climate and habitability on early Mars. Fluvial runoff erosion would require very different climatic conditions than those we observe today on Mars. Frequent precipitation and an active hydrological cycle (*10-14*) are necessary to support a significant amount of overland flow, with temperatures at least episodically reaching temperatures that allow liquid water to exist (*12-16*), with a thicker $CO_2$ atmosphere as a necessary factor.

Recent observations from spacecraft (*17, 18*) and from Mars meteorites (*19*) have provided evidence for long-lived habitable lakes and seas approximately 3.7 Ga on Mars, when most valley networks were formed. However, it remains unclear if these lakes and seas were fed by rainwater or snowmelt. Quantitative reconstruction of Mars paleoclimate is hampered by a lack of geologic constraints.

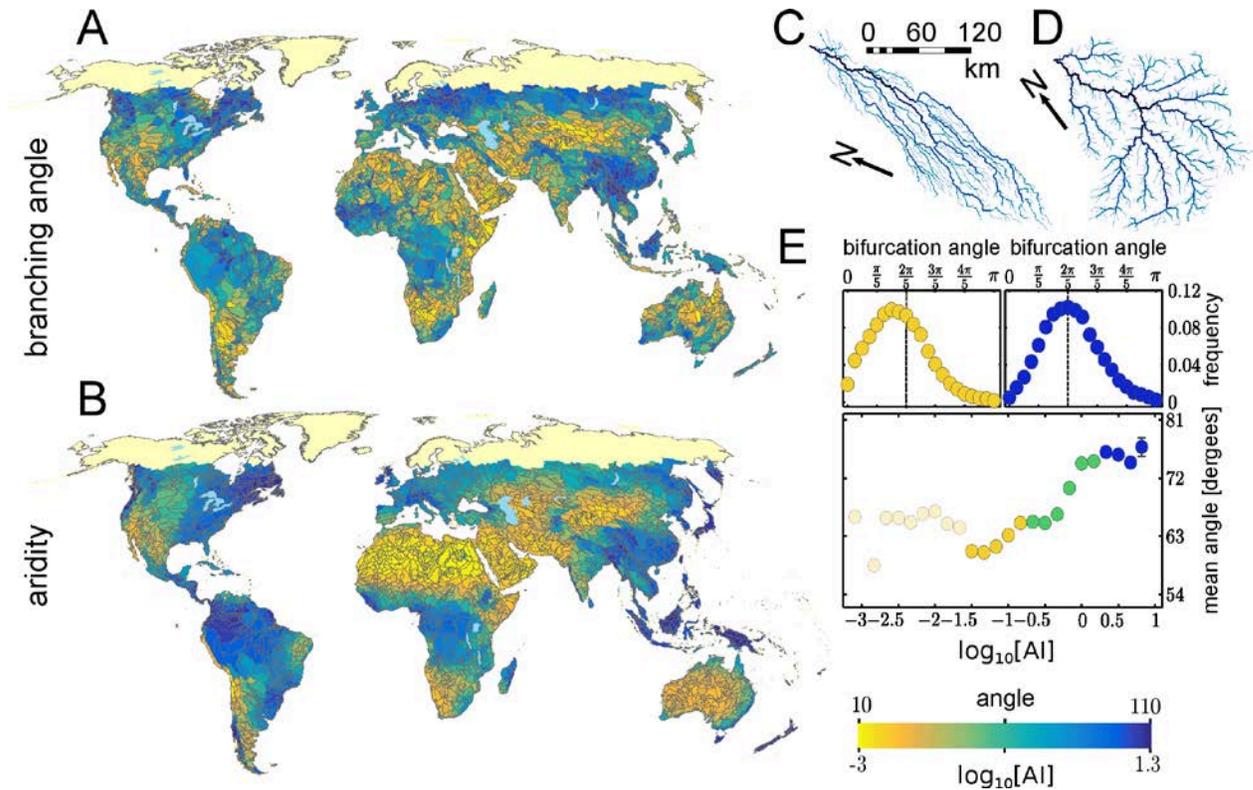

**Figure 1:** Global distribution of mean river network branching angles (A) and aridity index *AI* (B) averaged over HydroSHEDS' level 4 basins. Narrow branching angles are more likely to occur in arid regions (low *AI*, yellow), while in humid regions branching angles are usually wider (high *AI*, blue). Latitudes north of 50°N where no stream data are available are marked in cream. Panels (C) and (D) show examples of stream networks from arid (C) and humid landscapes (D). The arid network is located in eastern Algeria and the humid network lies in the Amazon rainforest at the border between French Guiana and Suriname. Mean branching angles for binned ranges of aridity index are shown in panel (E), together with the corresponding branching angle histograms for the arid (yellow) and humid (blue) tails respectively. The points with $\log_{10}(AI)<-1.5$ are shaded because they constitute only a small fraction (<5%) of the whole dataset spread over a wide range of aridity values; their inclusion or exclusion has no visual detectable effect on the global branching angle distribution.

Here we explore how Mars' channel networks can help constrain its climatic history. Recently, two independent studies using high-resolution channel networks across the contiguous United States have shown that branching angles of stream networks reflect the relative dominance in the underlying channel forming processes (*20, 21*) and their climatic drivers (*20*). Stream networks tend to branch at narrower angles in arid climates, where flash floods and overland flow are more common, while humid landscapes are more diffusively dissected, resulting in wider branching angles (*20*). Figure 1 shows that this relationship between stream network geometry and climatic controls is also observed in coarser-scale global maps (for details see supplementary information). Branching angles can easily be mapped from satellite images and are well preserved over time, thus providing a useful tool to infer channelization processes and their climatic drivers on extraterrestrial planets.

We now turn our attention to Mars' valley networks, derived from two independent high-resolution global datasets (Fig. 2A). Hynek & Hoke (*22*) manually extracted stream networks from infrared images with a pixel size of ~230m, and Luo & Stepinski (*23*) used automated extraction techniques to define channel networks from gridded laser altimeter data with a spatial resolution of about 128 pixels per degree, corresponding to roughly 463 meters at the equator. Two example networks from (*22*) are shown in Fig. 2C, D. These networks show a narrowly "feathered" geometry, similar to networks formed by overland flows in arid regions on Earth.

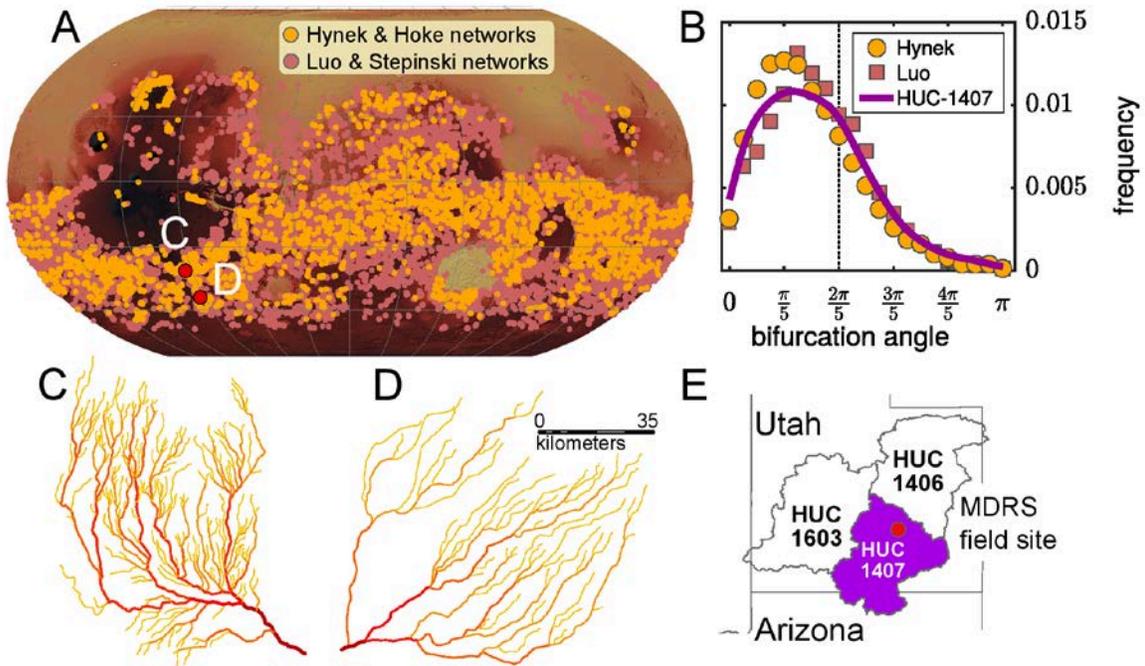

**Figure 2:** (A) outlet locations of the valley networks mapped by Hynek & Hoke (orange) (*22*) and Luo & Stepinski (rose) (*23*). Background shading indicates elevation. The corresponding branching angle distributions are shown in panel (B). The violet solid line represents the branching angle distribution in the Lower Green River, a basin in the arid southwestern United States. The modes of the three datasets are 36° for the Hynek & Hoke networks, 45° for the Luo & Stepinski networks, and 41° for the Lower Green River basin. These values are considerably smaller than the theoretical angle of $2\pi/5=72°$ (*24*) expected for groundwater-driven network growth (black dashed line). Two sample valley networks on Mars are shown in (C) and (D). Scale bar corresponds to both sites. (E) Map of the Lower Green River basin (HUC-1407), where the Mars Desert Research Station (MDRS, red circle) is located.

The global surveys of network branching angles on Mars confirm the generality of this observation. Figure 2B shows the branching angle statistics for the Hynek & Hoke (*22*) and Luo & Stepinski (*23*) network datasets, and although the two sets of networks have been extracted independently using different techniques, their branching angle distributions differ only in details, both peaking around $40°\pm5°$, (modes calculated by kernel density smoothing). These narrow characteristic angles suggest that Mars' valley networks may follow the regional topographic slope more closely than typical networks on Earth (*25*), which are often embedded into a smoothly varying landscape of valleys and hilltops, especially in humid regions.

The solid line in Fig. 2B compares the branching angles of the Martian networks with those of the streams of the Lower Green River basin in the arid southwestern United States (HUC-1407, Fig.2E) as mapped by the NHDPlusV2 dataset (supplementary material). This basin and its surroundings are thought to display similar landscapes to Mars (*26, 27*), and astronauts train there at the Mars Desert Research Station (MDRS) in anticipation of future Mars missions. Even though the mapping resolutions of river networks differ by almost a factor of ten between Mars and the desert southwest US, their branching statistics are strikingly similar. The mode of the branching angle distribution of the Upper Green River basin is approximately 41°, consistent with the global distribution on Mars. To check that this is not just a coincidence we also analyzed the branching angles of two neighboring basins (Fig. S2, supplementary material), which peak around 36° and 34°, respectively. Junction angles also vary with slope (*20*), but all of the networks shown in Fig. 2B have broadly similar slope distributions (Fig. S3, supplementary material), suggesting that slope differences are unlikely to be masking aridity differences between the Mars networks and the Earth analogues. The characteristic junction angles observed on Mars and in the desert southwest are significantly narrower than the characteristic branching angle of $2\pi/5=72°$ that is expected to arise from network growth driven by groundwater sapping (*24*).

These observations lead to the interpretation that Mars' channel networks were mainly formed by intense surface runoff driven by episodic precipitation events in an arid climate. This hypothesis is also supported by observations from drainage basin morphology (*26*) and erosional models (*27, 28*). A dry continental climate in the low- to mid-southern latitudes, where most of the channel networks reside, is consistent with the hypothesized ocean covering most of Mars' northern hemisphere (*29, 30*). This hemispheric segregation would imply an increasingly arid climate towards the south (*10, 31*).

To rule out the possibility that narrowly branched drainage patterns could also result from channelization in permafrost landscapes, we analyzed the high-resolution NHD drainage networks for the state of Alaska (supplementary materials). Figure 3 shows the branching angle statistics separated into regions with continuous, intermittent, and sporadic permafrost (violet, magenta and green respectively), where we pruned all first-order streams to achieve a comparable resolution to the NHDPlusV2 data for the rest of the continental US (see also Fig. S4, supplementary material). All three groups of networks have branching angle distributions that peak close to 72°, which is significantly wider than the branching angle distributions observed on Mars.

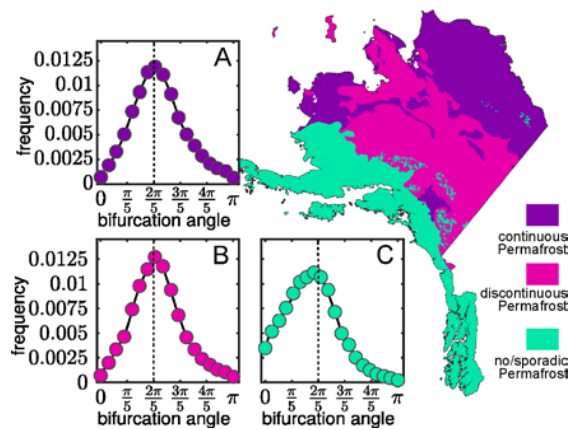

**Figure 3:** River network branching angles in the State of Alaska, separated into regions with continuous permafrost (violet, A), discontinuous permafrost (magenta, B), and absent or only sporadic permafrost (green, C). In all three cases, the bifurcation angle histograms peak at roughly 72° (dashed line), similar to the branching angles observed in other humid landscapes. The black solid lines in panels A-C show the kernel-smoothing density estimates of the branching angle distributions.

Most valley networks on Mars are thought to have been formed in a rather short epoch during the late Noachian and early Hesperian (32, 33). Lake coverage and valley incision depth both suggest a climatic optimum (11, 12, 32, 34) during this epoch. While many smaller tributaries may have been erased over time, and channels on valley floors are only rarely preserved, the planform branching pattern is probably the least-altered geomorphic feature of these rivers, in some cases even surviving channel inversion (*13*). The correlation of branching angles with climatic controls supports the recent shift from groundwater-dominated theories for Martian channel formation (*5-7*) to more recent precipitation-based theories (*10-12, 16, 33, 35*). Our analysis suggests that Mars' channel networks were formed in an arid continental climate with sporadic heavy rainfall events large enough to create surface runoff. Our results imply that the growth of Martian channel networks was dominated by near-surface flow, and that groundwater sapping, although regionally possible, played a relatively minor role.

## Acknowledgements
The authors thank W. Luo for providing his valley network dataset.

## Supplementary Materials:

Global aridity and branching angles

NHDPlusV2 data analysis for the arid southwest of the United States

NHD data analysis for the State of Alaska

Figures S1-S4

References (*36-41*)

**Supplementary Materials:**

**Global aridity and branching angles**

The branching angles in Fig. 1 were derived in the same way as in reference (20) using HydroSHEDS' (36) stream networks, following the schematic in Fig. S1. Branching angles are measured between the two orthogonal regression lines (black dashed lines) fitted to the two channel segments (blue) upstream of each junction, (Fig. S1). The aridity index $AI=P/PET$ was calculated based on precipitation $P$ and potential evaporation data $PET$ from WorldClim (37). For the maps in Fig 1A and Fig. 1B, the branching angles and aridity values were averaged over HydroSHEDS' level 4 drainage basins. The color scale runs from yellow for low branching angles and low $AI$ to blue for wide branching angles and high $AI$. The corresponding correlation function is shown in Fig. 1D where the branching angles were binned with respect to $\log_{10}(AI)$. Colors indicate arid (yellow, $AI < 0.2$ ), intermediate (green, $0.2 < AI < 2$) and humid (blue, $AI > 2$) climates (38). The histograms of the wet and dry tails are shown on top of the correlation curve and peak around 55° for dry regions and 72° for humid regions respectively. Two sample networks are shown in Fig. 1C to give a visual impression of the different network geometries in dry and wet regions.

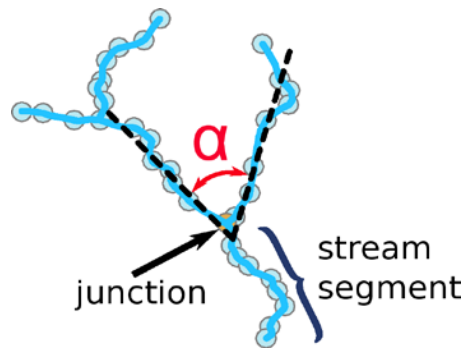

**Figure S1**: Measurement of the branching angle as defined in (20). Each stream segment (blue line) is defined by a series of points (blue circles). The dashed black lines indicate the orthogonal regression fits to the stream segments. The branching angle is measured between the regression lines of the two segments above each junction.

## NHDPlusV2z' data analysis

To compare Mars' valley networks with those of an arid landscape on Earth, we analyzed the stream networks in the arid southwest of the United States as mapped by the NHDPlus dataset in version 2 (*39*), (see Fig. 2B, Fig. S2 and Fig. S3).

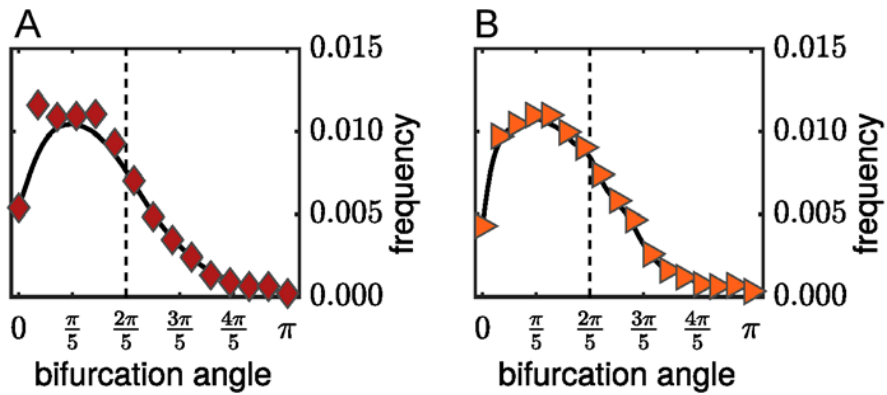

**Figure S2:** Histograms for branching angles derived from NHDPlusV2 data in two additional basins in the southwest of the United States, namely, (A) the Upper Colorado Dirty Devil basin, HUC-1406, and (B) the Escalante Desert-Sevier basin, HUC-1603. The black solid line corresponds to the kernel density smoothed distribution and the dashed line marks the theoretical bifurcation angle of groundwater driven growth(*24*).

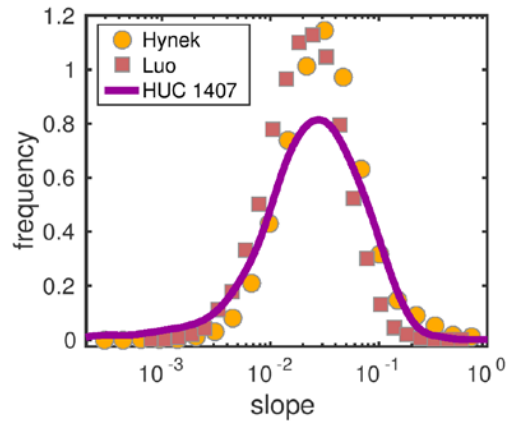

**Figure S3:** Histograms of valley slope for the two Mars datasets (Hynek & Hoke, orange circles, Luo & Stepinski, rose squares) and the Lower Green River basin (HUC-1407, violet solid line) as mapped by the NHDPlusV2 dataset.

## NHD data analysis for the State of Alaska

Because NHDPlus data are not available for the State of Alaska, we performed our branching angle analysis for Alaska's the permafrost landscapes based on the full NHD data provided by USGS (*40*). Horton-Strahler orders were calculated using the ArcGIS tool RivEX (*41*).

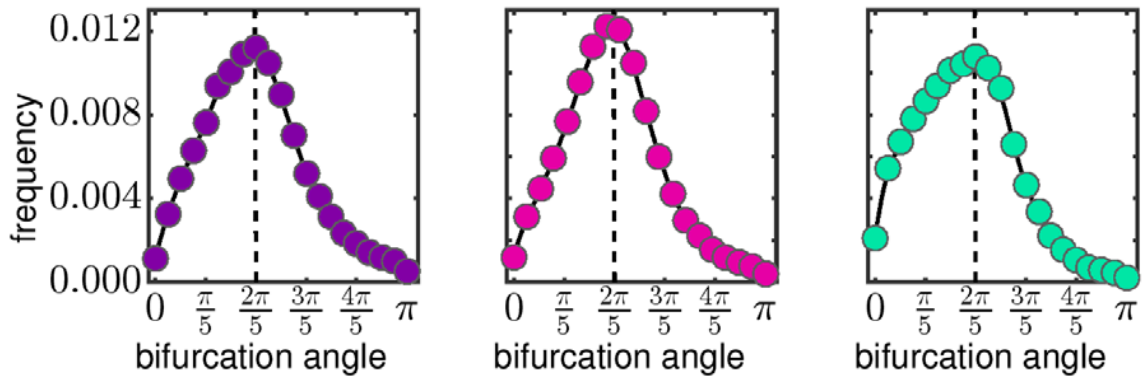

**Figure S4:** Branching statistics of the raw NHD streams of the State of Alaska (*40*) categorized in regions with continuous permafrost (violet), discontinuous permafrost (magenta), and no permafrost (green). The dashed line marks the theoretical bifurcation angle of groundwater-driven growth. In this figure, in contrast to Fig. 3 in the main paper, first-order channels have not been pruned. Comparing the two figures shows that pruning the first-order channels has a negligible effect on the junction angle distributions. The black solid line corresponds to the kernel density smoothed distribution and the dashed line marks the theoretical bifurcation angle of groundwater-driven growth.